\documentclass[a4paper,12pt]{article}
\usepackage{amsmath}
\usepackage{graphicx}
 \usepackage{amsthm, amssymb, amsfonts}
\usepackage{float}
\usepackage{textcomp}
\usepackage{latexsym}
\usepackage{amsfonts}
\usepackage{fancyhdr}
\usepackage{footnote}
\usepackage{fancyheadings}
\author{Timothy Chibueze${^1}$ and Ranjan Chaudhury${^2}$}
\date{}
\title{Synthesis of the Conventional Phenomenological Theories of Superconductivity with Marginal Fermi Liquid Model}

\begin{document}
%\footnote{Timothy}
%\end{document}
%\pagestyle{fancy}
%\twocolumn[
 %\begin{twocolumnfalse}
\maketitle
\footnote{${}$timothy.chibueze@unn.edu.ng, Department of Physics and Astronomy, University of Nigeria Nsukka, Nigeria; ${^2}$ranjan@bose.res.in, S.N Bose National Center for Basic Sciences Kolkata, India; 
}
\begin{abstract}
In this work we have done phenomenology based model calculations for some of the thermodynamic and electrodynamic
properties of the strongly correlated superconductors of Cuprate type. The method involves the application of the theoretical
result for electronic specific heat in the normal phase from Marginal Fermi Liquid theory
%using a direct approach similar to the Fermi liquid (FL) and phonon-like treatment obtained by Kuroda and Varma} 
to the Gorter-Casimir two fluid model to derive the temperature dependence of the critical magnetic field corresponding to a
type-I system, using the standard variational technique.
 We also applied this modified two fluid scheme to the London theory and obtained an expression 
for the temperature dependence of the magnetic field penetration depth in the superconducting phase. 
Our results are in fairly good agreement with other theoretical results based on different approaches, as well as with the experimental results. 
\end{abstract}
%\tableofcontents
%\end{twocolumnfalse}
 %]
\section{Introduction}
 High temperature superconductivity in Cuprates has taken the centre stage of modern condensed matter physics since its discovery
 in 1987 because of the unusual normal state properties of these materials combined with the very rich phase diagram, besides 
the superconducting transition temperatures in the range of 40K-164K. These systems exhibit deviations from the Fermi 
liquid phenomenology in large regime of stoichiometric compositions. Moreover, the conventional microscopic theory is not always
 successful to explain the properties in the superconducting phase satisfactorily.
 On a phenomenological level, the behaviour below the optimal 
doping in the normal phase seems to display `marginal Fermi liquid' (MFL) behaviour in the normal phase [19].\\
 One of the most important features observed in experiments in the normal phase of the cuprates is the linear temperature 
dependence of dc resistivity, which 
below the optimal doping persists in an enormous temperature range from a few kelvin to much above room temperature.\\
The studies of the electrodynamic properties in the superconducting phase provide a clear
 phenomenological scenario, reveal information regarding the pairing state, the energy gap and the electronic density of
 states and thus provide important indications on the mechanism of high temperature superconductivity.
 A phenomenological model describing  the marginal Fermi liquid behaviour
of cuprates has been put forward by Varma and co-workers but its microscopic origin remains highly controversial. 
To our knowledge, no microscopic theory has so far been able to provide a satisfactory explanation for the phenomenon of high 
temperature superconductivity and anomalous normal phase properties of cuprates despite tremendous efforts during 
the last 3 decades [3].\\

\begin{figure}[H]
\begin{center}
\includegraphics[width=0.6\textwidth]{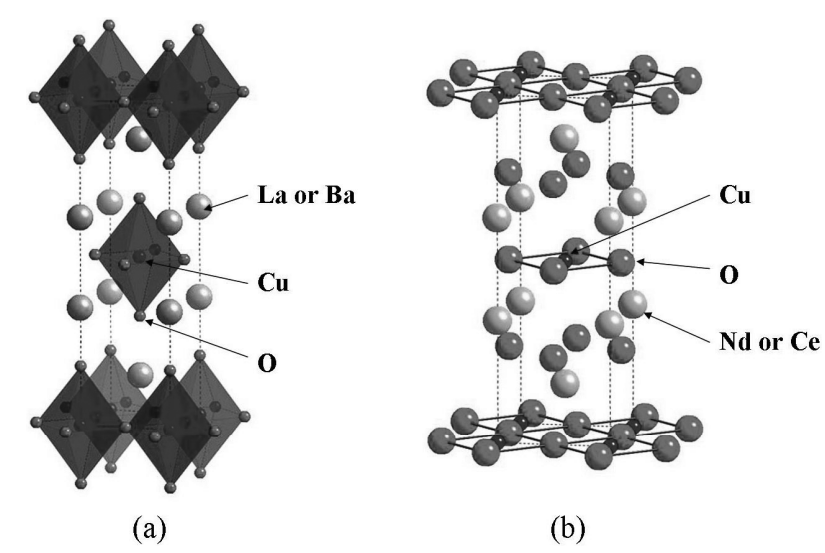} 
\caption{Crystal structures of (a) $La_{2-x}Ba_xCuO_4$ and (b) $Nd_{2-x}Ce-xCuO_4$ superconductors}
\end{center}
\end{figure}
\section{Gorter-Casimir Two Fluid Model}
Gorter-Casimir two fluid model is an 'ad hoc' model and it is based on two fundamental assumptions:\\
\begin{enumerate}

\item The superconducting state of a system is made up of two kinds of species viz. super-electrons and normal electrons.
 The perfectly ordered state occurs only at zero temperature and consists of super-electrons only and
\item The order parameter associated with the superconducting state is proportional to the number density of super-electrons 
and is dependent on temperature.
\end{enumerate}

Let x represent the volume fraction of electrons belonging to the normal fluid and $1-x$, that belonging to the superfluid.
 Gorter and Casimir assumed the following formal analytical form for the free energy density of the electrons [22]:\\
\begin{equation}\label{1}
 F_S(x,T)=x^{\frac{1}{2}}f_n(T)+(1-x)f_s(T)
\end{equation}

where\\
\begin{equation}
 f_n(T)=-\frac{\gamma}{2}T^2
\end{equation}
is the free energy density of the normal fluid
and\\
\begin{equation}
 f_s(T)=-\beta
\end{equation}
 is that for superfluid with
$\gamma$ being the Sommerfeld constant and it is proportional to the single electron density of states per unit volume 
$N(0)$ at the Fermi and $\beta$ is an unknown parameter to be determined later.
Minimizing the free energy density function $F(x,T)$ with respect to variations in $x$, one finds the equilibrium fraction 
of normal electrons at a temperature T.\\
\begin{equation}
\frac{\partial F(x,T)}{\partial x}=\frac{1}{2}x^{-{\frac{1}{2}}}f_n(T)-f_s(T)=0
\end{equation}
At $T=T_C$, $x=1$ \\Thus we have\\
\begin{equation}
 F_n=-2\beta\left(\frac{T}{T_c}\right)^2
\end{equation}

and\\
\begin{equation}
 F_n(T)-F_s(T)=\beta \left(1-\left(\frac{T}{T_c}\right)^2\right)^2
\end{equation}
From the thermodynamic relation, it can be shown that 
\begin{equation}
 \frac{H_c^2 (T)}{8 \pi}= F_N (T)-F_s (T)
\end{equation}
where $\frac{H_c^2 (T)}{8 \pi}$ is the stabilization (condensation) energy density of the pure superconducting state and $H_c \ (T)$ is 
the critical magnetic field.
 This leads to\\
\begin{equation}
 H_c(T)=H_o\left(1-\left(\frac{T}{T_c}\right)^2\right)
\end{equation}
where $H_o$ is the the critical magnetic field at zero temperature [6,7].\\

\section{The London Theory}

The brothers, 
H. London and F. London in 1935 [13] gave a phenomenological description of the electrodynamic properties of superconductors 
by proposing a scheme 
based on a two fluid type concept with super fluid and normal fluid densities $n_s$ and $n_n$ associated with 
velocities $v_s$ and $v_n$ respectively.The zero frequency penetration depth is a measure of the distance scale 
on which a static magnetic field will penetrate into a superconductor.
Although the superconductor in the bulk has the property that it excludes all the magnetic flux, because of the 
superconducting screening current, it is in the surface layer that the field may still penetrate [19].\\
The first London equation is
\begin{equation}
 \frac{\partial\bar{J}_s}{\partial t}=\frac{n_se^2E}{m}
\end{equation}\\

The second London equation is
\begin{equation}
 \nabla\times\bar{J}_s=-\frac{n_se^2}{mc}B
\end{equation}

and the London penetration depth $\lambda_L$ is given as
\begin{equation}
 \lambda_L=\left(\frac{mc^2}{4\pi n_se^2}\right)^{\frac{1}{2}}
\end{equation}

\section{The BCS Theory}

 Bardeen, Cooper and Schrieffer (BCS) proposed a microscopic Hamiltonian
for a superconductor, which is based on the idea of Cooper pairing [25]. Using
this theory, they were able to successfully describe the interaction between electrons forming Cooper pair.

The BCS theory has a parameter $g$ defined as
\begin{equation}
 g=N(o)V_{eff}
\end{equation}
where $V_{eff}$ is the magnitude of the effective attractive interaction between the electrons forming a Cooper pair.
From BCS equation in the weak coupling regime, one has the following equation for $T_c$ 
\begin{equation}
 T_c=1.13\theta_c\exp \left(-\frac{1}{g}\right)
\end{equation}
where $\theta_c$ is the temperature equivalent of the characteristic energy of the bosonic excitation mediating the pairing
interaction.
%{and it is equivalent to the characteristic frequency in the marginal Fermi liquid theory.}
In the weak coupling regime, $0<g<0.25$.\\

Hereafter we would assume equation (13) to be valid even when the pairing is mediated by high energy electronic boson.

\section{A Phenomenological Marginal Fermi Liquid Theory}
In general, the unusual normal state properties of the high temperature superconducting copper-oxide compounds indicate a 
scattering rate for the itinerant electrons, that is linear in frequency $\omega$ and linear in temperature T over a large region. 
This implies that these materials can not satisfactorily be described by the conventional Fermi liquid picture. 

Varma et al [20,22] postulated that in the copper oxide system, there are charge and spin density fluctuations of the electronic 
system, which are significantly distinct from those in the conventional Fermi Liquid. These two excitations however have similar
 behaviour. These fluctuations lead to a new contribution to the polarisability of the electronic medium that would renormalize 
the electron through the self energy in accordance with the observed scattering rates.\\
Their proposal for this contribution to the polarisability is as follows:\\
\begin{equation}
 Im \ P(q,w)=\Bigg\{^{-N(0)\frac{w}{T}, \qquad for \qquad |w|<T}_{-N(0)sign w \qquad for \qquad |w|>T}
\end{equation}
\\
where N(0) is the single particle density of states at the Fermi energy [1].

Kuroda and Varma [3] calculated the specific heat of the marginal Fermi liquid in the normal phase using a Fermi liquid-like formula
in the presence of electron-boson coupling constant. This boson is taken to be the itinerant  particle-hole pair (exciton) itself
 in the normal
 state.  They obtained the electronic specific heat $C_v$ of the marginal Fermi liquid as
 
\begin{equation}
 C_v=N(0)\left(3+2\ln\frac{\theta_c}{T}\right)T
\end{equation}
where $\theta_c$ is the characteristic temperature corresponding to the energy of the excitonic boson in the marginal Fermi liquid theory 
and assuming coupling coefficient $\lambda_{+}$=1.
 
%\begin{document}
\section{Synthesis of Gorter-Casimir Two Fluid Model with Marginal Fermi Liquid Model}

The free energy of conduction electrons in a metal is given by
\begin{equation}
f_n=U-T \displaystyle\int_0^T \dfrac {C_v (T)}{T} dT 
\end{equation}\\
 where  $ \ U=\int_0^T C_v(T) dT$
and represents the internal energy of the electrons in the system.\\
We can then extend the above result and make use of Equation (15) to arrive at the following expression for the free energy density 
of the electrons in the normal phase of the marginal Fermi liquid:\\
\begin{equation}
f_n=-N(0)\left(3+\ln\frac{\theta_c}{T}\right)T^2
\end{equation}
Making use of the two fluid model [see Equation (1)], the total electronic free energy density in the superconducting phase 
of the marginal Fermi liquid is  now given as\\
\begin{equation}\label{2}
F_S(x,T)=-x^{\frac{1}{2}}N(0)\left(3+\ln\frac{\theta_c}{T}\right)T^2\\ +(1-x)(-\beta)
\end{equation}
\begin{equation}
\frac{\partial F(x,T)}{\partial x}=-\frac{1}{2}x^{-\frac{1}{2}}N(0)\left(3+\ln\frac{\theta_c}{T}\right)T^2\\+\beta=0
\end{equation}
This leads to the following equation after incorporating the expression for $\beta$ determined from the condition that for $T$ approaching
$T_c$, $x$ approaches 1,
\begin{equation}\label{3}
x=\left(\frac{3+\ln\frac{\theta_c}{T}}{3+\ln\frac{\theta_c}{T_c}}\right)^2\left(\frac{T}{T_c}\right)^4
\end{equation}
where now $x$ represents the fraction of electrons in the normal fluid existing in the form of the marginal Fermi liquid.\\
Substituting the expression for $x$ from equation (20) into (18) gives\\

\begin{equation}
F_S(T)=-\frac{1}{2}N(0)\\ \times \left[\frac{\left(3+\ln\frac{\theta_c}{T}\right)^2}{\left(3+\ln\frac{\theta_c}{T_c}\right)}\frac{T^4}{{T_c}^2}+\left(3+\ln\frac{\theta_c}{T_c}\right)T_c^2 \right]
\end{equation}\\
Since $F_N(T)$=$f_n(T)$, it is given by equation (17) itself.\\
We recall that
$$C_v^S=-T\frac{\partial^2F_s(T)}{\partial T^2}$$\\
This yields,\\
\begin{equation}
C_v^S= N(0)\frac{T^3}{T_c^2}\frac{1}{\left(3+\ln\frac{\theta_c}{T_c}\right)}\\ \times\left[6\left(3+\ln\frac{\theta_c}{T}\right)^2-7\left(3+\ln\frac{\theta_c}{T}\right)+1\right]
\end{equation}
\begin{figure}[H]
\begin{center}
\includegraphics[width=0.6\textwidth]{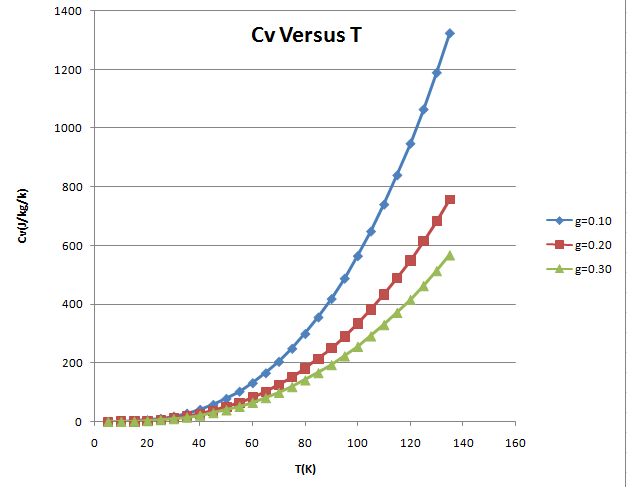}
\caption{Graph of specific heat in the superconducting phase $C_v$ against temperature T.}
\end{center}
\end{figure}

Also\\
\begin{equation}
C_v^N=N(0)\left(3+2\ln\frac{\theta_c}{T}\right)T
\end{equation}
The difference between the specific heat capacity in the superconducting state $C_v^S$ and the specific heat capacity in the normal
 state $C_v^N$, called the specific heat jump $\triangle C_v$ is given as\\ 
$$\triangle C_v=C_v^S-C_v^N$$\\

At the critical temperature $\triangle C_v$ is given as\\
\begin{equation}
 \triangle C_v|_{T=T_c}= N(0)T_c\frac{1}{\left(3+\ln\frac{\theta_c}{T_c}\right)}\\ \times \left[6\left(3+\ln\frac{\theta_c}{T_c}\right)-7+\frac{1}{\left(3+\ln\frac{\theta_c}{T_c}\right)}\right]\\ -N(0)\left(3+2\ln\frac{\theta_c}{T_c}\right)T_c 
\end{equation}

At critical temperature $T_c$, the ratio of the two types of specific heat is given as\\
\begin{equation}
\frac{C_v^s}{C_v^N}\bigg|_{T=T_c}= \frac{1}{\left(3+2\ln\frac{\theta_c}{T_c}\right)}\\ \times \left[6\left(3+\ln\frac{\theta_c}{T_c}\right)-7+\frac{1}{\left(3+\ln\frac{\theta_c}{T_c}\right)}\right]\\
\end{equation}

The normalized specific heat jump at the transition temperature is given as\\
$$R=\frac{\triangle C_v}{C_v^N}\bigg|_{T=T_c}$$\\
and we have\\
\begin{equation}
 R= \frac{1}{\left(3+2\ln\frac{\theta_c}{T_c}\right)}\\ \times \left[6\left(3+\ln\frac{\theta_c}{T_c}\right)-7+\frac{1}{\left(3+\ln\frac{\theta_c}{T_c}\right)}\right]-1
\end{equation}
\begin{figure}[H]
\begin{center}
\includegraphics[width=0.6\textwidth]{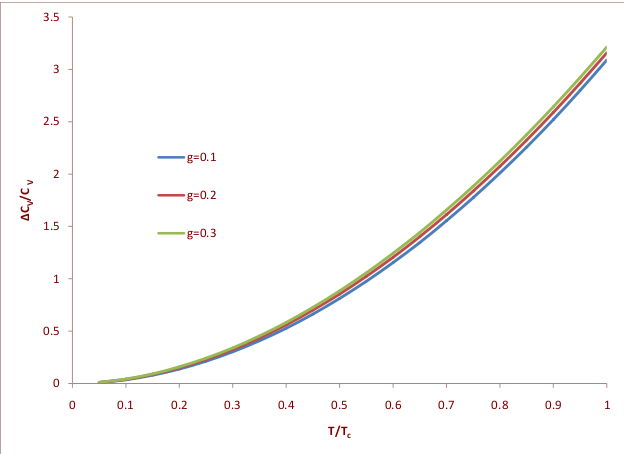}
\caption{Graph of normalized specific heat jump in the superconducting phase $C_v$ against temperature $\frac{T}{T_c}$}
\end{center}
\end{figure}

At transition temperature, the normalized slope of the specific heat jump is given as\\
$$D=\frac{\frac{d(\triangle C_v)}{dT}}{\frac{dC_v^N}{dT}}\bigg|_{T=T_c}$$\\
\begin{equation}
D=\frac{1}{\left(1+2\ln\frac{\theta_c}{T_c}\right)}\\ \times \left[18\left(3+\ln\frac{\theta_c}{T_c}\right)-33+\frac{10}{\left(3+\ln\frac{\theta_c}{T_c}\right)}\right]-1
\end{equation}
\begin{figure}[H]
\begin{center}
\includegraphics[width=0.8\textwidth]{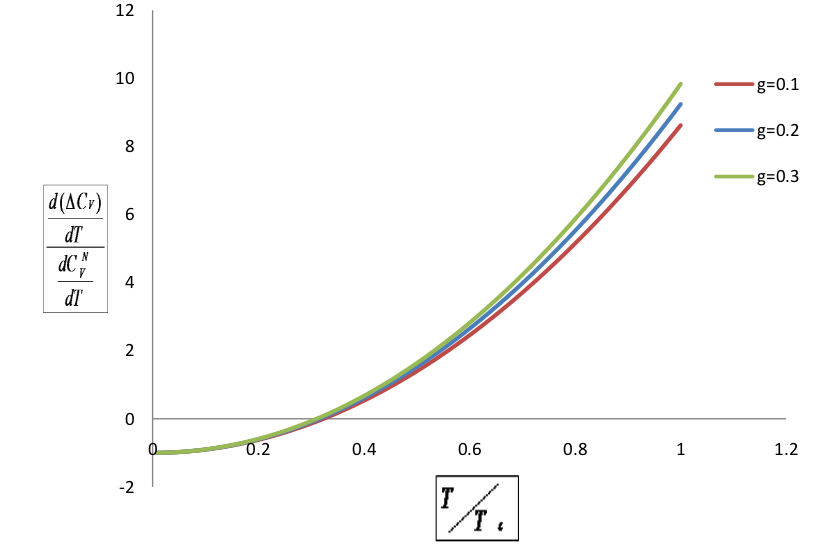}
\caption{Graph of the normalized slope of specific heat jump in the superconducting phase $\Delta C_v$ against normalised temperature $\frac{T}{T_c}$}
\end{center}
\end{figure}
\begin{figure}[H]
\begin{center}
\includegraphics[width=0.8\textwidth]{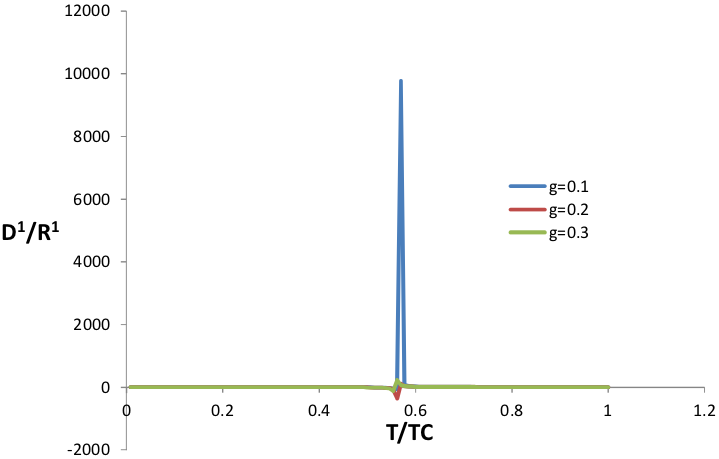}
\caption{Graph of ratio of the normalized slope of  specific heat jump and normalized specific heat
 jump in the superconducting phase $\frac{D}{R}$ against normalised temperature $\frac{T}{T_c}$}
\end{center}
\end{figure}

Making use of equations (17) and (23) in equation (7), we have 

\begin{equation}
\frac{{H_c}^2(T)}{8\pi}=\beta\left[1-\left(\frac{3+\ln\frac{\theta_c}{T}}{3+\ln\frac{\theta_c}{T_c}}\right)\left(\frac{T}{T_c}\right)^2\right]^2
\end{equation}
At $T=0$, $H_c(T)=H_1$ and we have\\
\begin{equation}
 \frac{H_1^2}{8\pi}=\beta
\end{equation}

and\\

\begin{equation}
H_c(T)=H_1\left[1-\left(\frac{3+\ln\frac{\theta_c}{T}}{3+\ln\frac{\theta_c}{T_c}}\right)\left(\frac{T}{T_c}\right)^2\right]
\end{equation}
This is a departure from the conventional Two Fluid Model behaviour expected on the basis of the normal state modelled as a Fermi Liquid.
From equation (13) we have

 $$ln\left(\frac{\theta_c}{T_c}\right)=\frac{1}{g}-0.12$$

Equation (30) gives the expression for the temperature dependence of the critical magnetic for a superconductor arising from the marginal
Fermi liquid normal phase.
\begin{figure}[H]
\begin{center}
\includegraphics[width=0.6\textwidth]{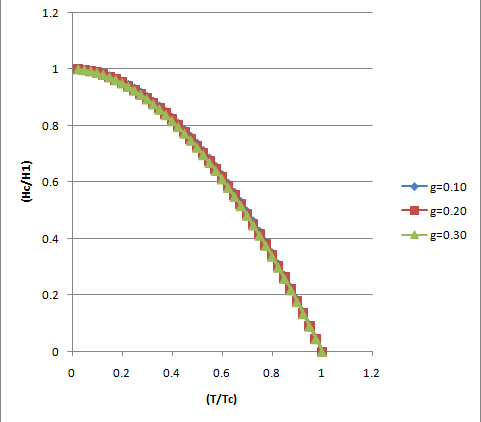}
\caption{Graph $\frac{H_c}{H_1}$ against $\frac{T_c}{T}$ from equations (34(a)) and (39).}
\end{center}
\end{figure}

\section{Synthesis of London Theory with MFL Model}

From the two fluid model,\\
\begin{equation}
 \frac{n_s}{n}=1-x
\end{equation}
and we have
\begin{equation}
 n_s=n\left[1-\left(\frac{3+\ln\frac{\theta_c}{T}}{3+\ln\frac{\theta_c}{T_c}}\right)^2\left(\frac{T}{T_c}\right)^4\right]
\end{equation}
where $n_s$ and $n$ retain their meaning.\\
From London's equations,\\
\begin{equation}
 \lambda_L(T)=\left(\frac{mc^2}{4\pi n_se^2}\right)^{\frac{1}{2}}
\end{equation}
where $c$ is a constant and $m$ and $e$ are the mass and charge of electron\\ respectively.\\
At $T=0$, $n_s=n$ and the penetration depth $\lambda_L(T)$ becomes $\lambda_L(0)$ and we get\\
\begin{equation}
 \lambda_L(0)=\left(\frac{mc^2}{4\pi ne^2}\right)^{\frac{1}{2}}
\end{equation}
Substituting for $n_s$ in equation (32) into equation (33) gives\\

\begin{equation}
 \lambda_L(T)=\frac{\lambda_L(0)}{\left[1-\left(\frac{3+\ln\frac{\theta_c}{T}}{3+\ln\frac{\theta_c}{T_c}}\right)^2\left(\frac{T}{T_c}\right)^4\right]^\frac{1}{2}}
\end{equation}
where $\lambda_{L(0)}=\frac{c}{\omega_p}$. The temperature dependence shows departure from usual behaviour.
\begin{figure}[H]
\begin{center}
\includegraphics[width=0.6\textwidth]{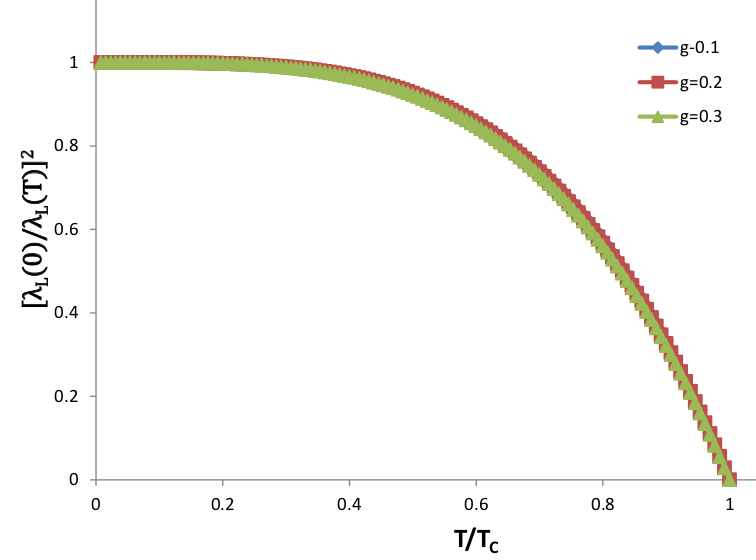}
\caption{Graph of $\left[\frac{\lambda_L(0)}{\lambda_L(T)}\right]^2$ against $\frac{T}{T_c}$}
\end{center}
\end{figure}

\section{Discussion of Results}
In our model, we have incorporated the normal phase properties described by marginal Fermi liquid theory into the structure of Gorter-Casimir two fluid model. 
Specific heat measurements 
give information on the electron-boson coupling strength. The BCS theory and its subsequent refinements based on the Eliashberg equations 
show that high critical temperatures in superconductors are favoured by high values of the frequencies of the bosons mediating
the pairing interaction and by the large electronic density of states at the Fermi level.\\

The quantity of interest is the difference between the electronic specific heats in the normal and superconducting phases.
 Our calculation shows that the normalized specific heat 
jump differs appreciably from $1.43$, the value corresponding to the BCS weak coupling limit for a superconducting transition from the 
conventional Fermi liquid phase.\\

At low temperatures, the lattice contribution to the total specific heat is small and can be accurately subtracted to extract the purely
electronic contribution. 
The normal phase specific 
heat can be obtained by applying a magnetic field of sufficient strength to cause the sample to become normal.\\
The ratio of the normalized slope of  specific heat jump and normalized specific heat
 jump in the superconducting phase $\frac{D}{R}$ at critical temperature
 $T_c$ is $4.1197$, $4.2616$, and $4.4110$ for $g=0.1, \ 0.2 \ and \ 0.3$ respectively.\\  
In the oxide superconductors, there are difficulties associated with these measurements. Because the superconducting critical temperatures of 
these oxide materials 
are relatively high, the lattice contribution to the total specific heat is quite large compared to the electronic contribution.
 An additional 
complication is that it is only possible to get normal state data close to critical temperature as the critical fields are quite large 
and are difficult to produce in the laboratory.\\

Figure (7) represents the experimental results for specific heat corresponding to YBCO. Comparing with figure $2$, observe that at low temperatures, there is an upturn 
in the specific heat rather than the expected 
exponential decay. However, there is still a linear term but there is no consensus yet on its origin. Analysis of the experimental data 
is usually done by assuming that the BCS relation $\frac{\triangle C}{\gamma T_c}= 1.43$ holds. However it is pointed out by Beckman 
et al [27] 
that $\gamma$ extracted by this analysis is not in good agreement with values from high $T_c$ magnetization experiments and band 
structure 
calculations.\\

\begin{figure}[H]
\begin{center}
\includegraphics[width=0.6\textwidth]{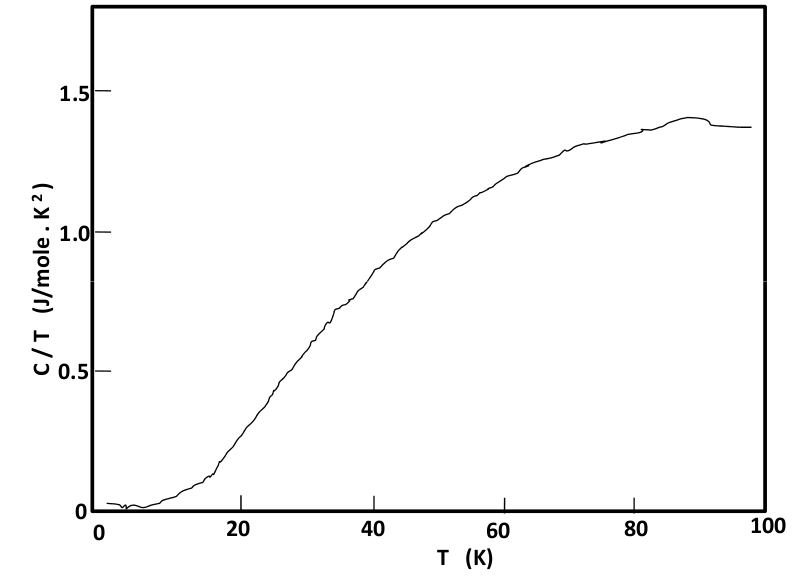}
\caption{Experimental result for the specific heat of $YBa_2Cu_3O_7$.}
\end{center}
\end{figure} 

Loram and Mirza [17] have used differential calorimetry on YBCO samples and report a normalized specific heat jump of $4.1$. 
Philips at al have reported a value of $4.8$.\\

From various observations, it would seem that there is a strong evidence for the specific heat jump to be large in the
 high $T_c$ materials.
This large value of the normalized specific heat jump is consistent with the result of $\sim 3.02$ in the model of synthesizing 
the Gorter-Casimir two fluid model with marginal 
Fermi liquid theory as done in the BCS weak coupling regime this thesis. \\
In the second part of this work, we calculated the magnetic field penetration depth by applying the MFL modified two fluid model. 
The main aim was to investigate the effect of the charge and spin density fluctuations of the electronic system in the copper
 oxide materials.
This we have done within the scope of the BCS weak coupling theory.
 \begin{figure}[H]
\begin{center}
\includegraphics[width=0.6\textwidth]{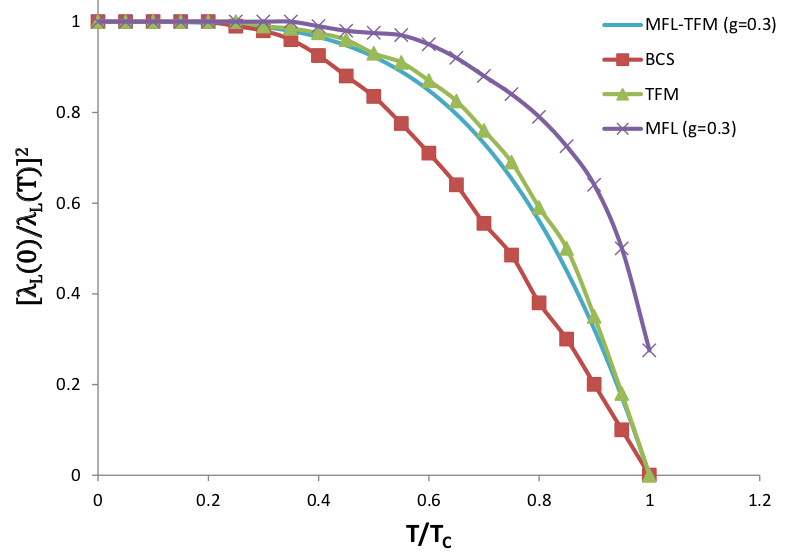}
\caption{Comparism of the results of penetration depth from the BCS, TFM, MFL, and our calculation (TFM-MFL)}
\end{center}
\end{figure} 
In figure (9) we have compared various results of the London penetration depth for the cuprate superconductor.
 The BCS weak coupling, the 
Gorter-Casimir two fluid model (TFM), the marginal Fermi liquid model (MFL) as done in the strong coupling regime by Nicol et al [19] 
and the synthesis of the MFL theory with London theory within the Two Fluid scheme (MFT-TFM) calculated in this piece of work.
In general muon spin relaxation ($\mu$SR) experiments tend to agree more with the two fluid model. The most resent experimental result
indicates temperature dependence conforming more to MFL-TFM like behaviour [28]\\
Note, however, that there is currently no consensus on the precise shape of $\frac{\lambda_L^2 (0)}{\lambda_L^2 (T)}$ in YBCO [15]

The London penetration depth from our result is close to the result of other results. If we extend our calculation to the BCS
 strong coupling 
regime, we hope to get a result closer to the experimental result.  
\section{Conclusion}
 In this research, we have applied the results from the marginal Fermi liquid theory to the $(i)$ Gorter-Casimir two fluid model 
and $(ii)$ London theory and 
used these to calculate some thermodynamic properties like the specific heat jump and the temperature dependence of the critical 
magnetic 
field. We also calculated the electrodynamic property in particular magnetic field penetration depth. The results of our calculations are closer to the 
experimental results obtained for Cuprates, than those from each of the phenomenological theories within the framework of ordinary Fermi liquid assumptions, independently.\\

In this study, we have only modified the normal fluid part of the Gorter-Casimir two fluid model. A more accurate result can be 
obtained 
by modifying the super fluid part as well. One method of doing this is to use a scheme based on many body formalism which leads
 to the free
 energy of the full superconducting phase for a MFL superconductor [3].  From this one can in principle subtract
 the normal fluid free energy density and thereby extract the super-fluid contribution corresponding to MFL.\\

Our methodology will be extended to a type-II superconducting system in future.
%\end{document}

\bibliographystyle
\cleardoublepage

\end{document}